\documentclass[preprint,12pt]{elsarticle}

\usepackage{graphics}

\usepackage{amssymb}
\usepackage{bm}

\usepackage{amsthm}

\usepackage{xcolor}

\begin{document}

\begin{frontmatter}



\title{Critical Current Anomaly at the Topological Quantum Phase Transition in a Majorana Josephson Junction}

\author[sysu]{Hong Huang}
\author[su]{Qi-Feng Liang}
\author[sysu]{Dao-Xin Yao \corref{cor2} }
\author[sysu]{Zhi Wang \corref{cor1}}
\cortext[cor1]{Corresponding author: Zhi Wang; Email: physicswangzhi@gmail.com; Tel: +86-20-84111107}
\cortext[cor2]{Corresponding author: Dao-Xin Yao; Email: yaodaox@mail.sysu.edu.cn; Tel: +86-20-84112078}

\address[sysu]{School of Physics, Sun Yat-sen University, Guangzhou 510275, China}
\address[su]{Department of Physics, Shaoxing University, Shaoxing 312000, China}

\begin{abstract}
Majorana bound states in topological Josephson junctions induce a $4\pi$ period current-phase relation. Direct detection of the $4\pi$ periodicity is complicated by the quasiparticle poisoning. We reveal that 
Majorana bound states are also signaled by the anomalous enhancement on the critical current of the junction. We show the landscape of the critical current for a nanowire Josephson junction under a varying Zeeman field, and reveal a sharp step feature at the topological quantum phase transition point, which comes from the anomalous enhancement of the critical current at the topological regime. 
In multi-band wires, the anomalous enhancement disappears for an even number of bands, where the Majorana bound states fuse into Andreev bound states. This anomalous critical current enhancement directly signals the existence of the Majorana bound states, and also provides a valid signature for the topological quantum phase transition.
\end{abstract}


\date{\today}
\end{frontmatter}

\section{Introduction}
The study of topological superconductivity has seen considerable progress since the theoretical proposal of implementing proximity effect to convert the conventional s-wave superconductivity into topological non-trivial p-wave superconductivity\cite{fu,sau,lutchyn,oreg,sato}. In general, topological superconductors contain all superconducting systems with a well defined non-zero topological number\cite{kanermp,zhangrmp}. In practice, however, topological superconductors often refer to those superconductors where the non-trivial topology give rise to a type of bizarre boundary states, the Majorana bound states (MBSs)\cite{read,kitaevwire,aliceareview,beenakkerreview,franzreview}.
MBSs are self-conjugate superconducting quasiparticles, which obey non-Abelian statistics\cite{ivanov,teo,aliceanphy}. Spatially separated MBSs define non-local topological qubits which are immune from local decoherence\cite{kanermp}. Braiding the MBSs acts as topological gates on these topological qubits\cite{kanermp,ivanov}. Therefore, MBSs are considered as a promising cornerstone for realistic quantum computation\cite{bkreview}, despite that these topological gates are not sufficient to make universal quantum operations.

MBSs have been reported in a number of experiments\cite{mourik,rokhinson,das,yazdani,albrecht,jia,kouwenhoven,molenkamp,moler,lv}, which mostly study the hybrid systems consisting of a conventional s-wave superconductor and a spin-orbit coupling nanowire\cite{mourik,rokhinson,albrecht}. The wire becomes topological superconducting with a combination of appropriate Zeeman energy, fine tunned chemical potential, strong spin-orbit coupling, and proximity induced Cooper pairing. Each end of the wire pins one MBS, which brings in unique transport signatures such as the resonant Andreev effect\cite{law}, the crossed Andreev effect\cite{beenakker08}, the quantized thermal conductance\cite{Akhmerov}, and the fractional Josephson effect\cite{kitaevwire,jiang}. The fractional Josephson effect draws particular interest since it is a phase coherent signal. It describes the coherent single electron tunneling through the MBS channel, which contributes a $4\pi$ period current-phase relation\cite{kitaevwire}. Direct measurement of this $4\pi$ period current-phase relation is hindered by the stringent requirement of complete elimination of the quasiparticle poisioning and MBS poisioning\cite{wangsr}.

Quasiparticle poisioning is the external single-electron tunneling into nanowire, causing  decoherence in the system. MBS poisioning is the annihilation of two MBSs causes by the strong interaction between them. Up to now, only indirect signals have been discovered experimentally in several systems\cite{rokhinson,kouwenhoven,molenkamp}. 

\begin{figure}[h]
\begin{centering}
\includegraphics[clip=true,width=0.5\textwidth]{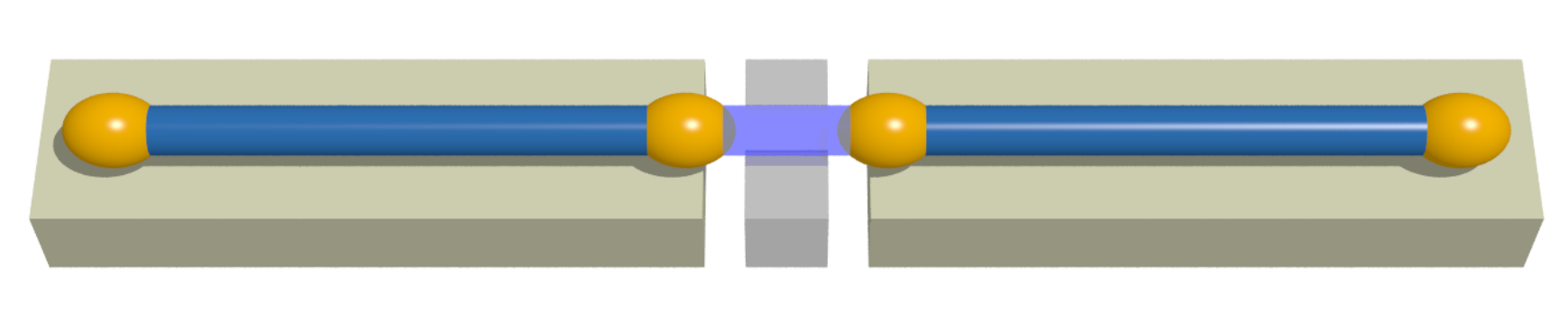}
\caption{(Color online). Schematic setup of a Majorana Josephson junction, with two conventional superconductors connected by a spin-orbit couping nanwire. A voltage gate is applied in the middle of the wire to produce a tunneling barrier. Four Majorana bound states $\gamma_{1,2,3,4}$ stay at the two sides of the tunneling barrier and the two ends of the wire.}
\label{F1}
\end{centering}
\end{figure}

Aside from the current-phase relation, Josephson junctions are also characterized by the critical current, which is the threshold of the phase driving current.
For a topological Josephson junction, the critical current consists of two parts: the $4\pi$ period component from the Majorana channel and the $2\pi$ period component carried by the quasiparticle channels\cite{beenakkerprb}. The former is small in comparison with the latter for bulk systems, since there is only one Majorana channel but many other quasiparticle channels\cite{grosfeld}.
However, the one-dimensional topological Josephson junction behaves differently when the electron tunneling is suppressed by a tunneling barrier. For these systems, the number of quasiparticle channels is restricted. In addition, each quasiparticle channel contributes only a small current for a high tunneling barrier, since it involves a second order perturbation process. In contrast, the Majorana channel contributes a larger Josephson current through a first order perturbation process.
Therefore, the MBSs should contribute a much larger critical current than the conventional quasiparticle channels for nanowire topological junctions.
If the wire is switched between topological and trivial states by a control parameter such as the Zeeman energy, we expect an anomalous sharp step for the critical current at the topological quantum phase transition (TQPT) point. The TQPT was predicted to be signaled by a quantized thermal conductance in Ref [29]. Here, we propose using this critical current anomaly as an alternative signal for the TQPT.

In this work, we show the existence of anomalous critical current enhancement in one-dimensional topological Josephson junctions. For this purpose, we adopt a nanowire hybrid Josephson junction as sketched in Fig. 1. The system consists of two conventional s-wave superconductors and a spin-orbit coupling nanowire. The wire is divided by a tunneling barrier which is produced by a voltage gate. This hybrid junction walks through a TQPT between the trivial phase and the topological superconducting phase when the Zeeman energy is increased from zero. We show that the critical current of the junction increases by orders when the system enters the topological phase, and exhibits a step-like feature at the TQPT point. We also study the behavior of the critical current when the Zeeman field is rotated, and show that the critical current anomaly also appears at the TQPT point. We further reveal that this step-like anomaly disappears for wires with an even number of sub-bands due to the fusion of the MBSs. 
These anomalous features for the critical current provide a signal for the TQPT and the existence of the MBSs, which must be helpful experimental detection since measuring the critical current is a routine procedure in experiments.

The rest of this paper is organized as follows. We present a toy model and the analytical results based on perturbation calculations in section II. Then we use Bogoliubov-de Gennes (BdG) approach to simulate the critical current of a realistic nanowire junction, and show the results for the increasing and rotating the Zeeman fields in section III. Afterwards we study the critical current for multi-layer systems in section IV. Finally, we give a summary in section V.

\section{Toy Model and Critical Current Enhancement}
The Majorana Josephson junction is consisting of a superconducting nanowire which is divided into two segments by a voltage gate\cite{mourik}. We view these two segments as two isolated wires, which are connected by the electron tunneling through the potential barrier. The minimal model for each segment is an one-dimensional chain with spin-orbit coupling, Zeeman energy, and superconducting pairing\cite{sarmamodel},
\begin{eqnarray}
H_\alpha&&= 
- t_\alpha  \sum_{ \langle i , j \rangle, \alpha, \sigma} c_{i , \alpha, \sigma}^\dagger c_{j , \alpha,\sigma}
 +\sum_{i, \alpha} \Delta_\alpha e^{i\phi_\alpha}  c_{i, \alpha,\uparrow}^\dag c_{i, \alpha,\downarrow}^\dag
\nonumber\\
&& +  \eta_\alpha  \sum_{i,  \alpha,\sigma, \sigma'}^{n}   c_{i+1,  \alpha,\sigma}^{\dag}   (i\sigma_y)_{\sigma\sigma^\prime}   c_{i , \alpha,\sigma'} 
- \mu_\alpha \sum_{ i ,  \alpha,\sigma} c_{i,  \alpha,\sigma}^\dagger c_{i, \alpha,\sigma}
\nonumber \\
&& + \sum_{i,  \alpha,\sigma\sigma^\prime}   c_{i, \alpha, \sigma}^\dag (V_x \sigma_x)_{\sigma\sigma^\prime}
c_{i, \alpha,\sigma'} , 
\end{eqnarray}
where $\alpha = L, R$ represents the left and the right segments of the wire, $\sigma = \uparrow, \downarrow$
represents the spin of the electron, $t$ is the nearest neighbor hopping in the tight-binding model, $\mu$ is the chemical potential, $\eta$ represents the spin-orbit coupling, $\Delta$ is the superconducting gap from the proximity effect, $\phi$ is the superconducting phase, and $V_{x}$ is the Zeeman energy from the horizontal Zeeman field. For simplicity, we consider identical parameter for the two segments $t_\alpha = t$, $\eta_\alpha = \eta$, $\Delta_\alpha = \Delta$, and $V_\alpha = V$. However, this does not change the physical results.
The left chain is connected to the right chain with an electron tunneling Hamiltonian,
\begin{eqnarray}
H_T = T \sum_{\sigma}   (c_{L, 0 ,\sigma}^{\dagger} c_{R,0,\sigma} +c^\dagger_{R,0,\sigma}c_{L, 0 ,\sigma}) .
\end{eqnarray}
where $T$ is the tunneling matrix between the two segments. To avoid quasiparticle poisioning and MBS poisioning, we assume that the system is in open boundary condition and the nanowire is long enough to seperate MBSs. In this paper, we assume each segment has 500 sites.

\begin{figure}[h]
\begin{centering}
\includegraphics[clip=true,width=0.5\columnwidth]{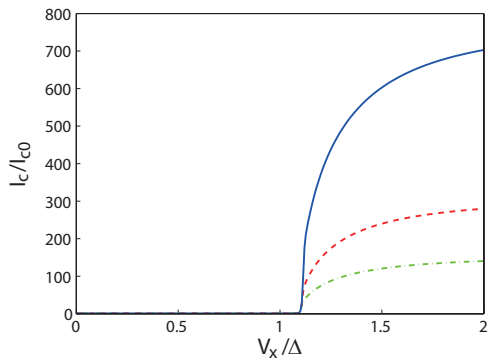}
\caption{(Color online). The critical current of the single-band Josephson junction in response to the Zeeman energy $V_x$. The solid, dashed, and dotted lines are for the case of tunneling matrix $T=0.002t$, $0.005t$, and $0.01t$, respectively. Each curve was normalized to the value of $V_x = 0$ for clarity. We choose the parameters in proportion to $\Delta$, other parameters are taken as $\mu= -43.6\Delta $, $t = 22\Delta$, $\eta=2.68\Delta$. Each current value in $V_x$=0 are $6.4\times10^{-7}\frac{e\Delta}{2\hbar}$, $4.0\times10^{-6}\frac{e\Delta}{2\hbar}$, $1.6\times10^{-5}\frac{e\Delta}{2\hbar}$, respectively.}
\label{F2}
\end{centering}
\end{figure}

The tunneling matrix is well controlled by the voltage gate. When the voltage gate creates a high potential barrier, the tunneling matrix $T$ becomes small and serves as a valid perturbation parameter. In the perturbation approach, two types of tunneling processes contribute to the Josephson current: the second order tunneling process through the quasiparticle channels, and the first order tunneling process through the MBSs\cite{aliceanphy}. We now make a qualitative estimation on the amplitude these two tunneling processes. We first examine the supercurrent from the quasiparticle channels. For this purpose, we take the trivial limit of zero spin-orbit coupling and Zeeman energy $\eta = V_x = 0$. Then the two segments become two simple one-dimensional s-wave superconductors. In this case, we can calculate the Josephson current with standard Green function technique (see appendix for details). The lowest order contribution to the current is second order in the tunneling matrix. We obtain a $2\pi$ period Josephson current $I = I_1 \sin \theta$, with $\theta= \phi_L - \phi_R$ the phase difference between the two chains. The critical current is
\begin{eqnarray}\label{conventionalcritical}
I_1 =   \frac{  e  \Delta T^2}{  2 (1 - \mu^2/ 4 t^2) \hbar t^2},
\end{eqnarray}
where $e$ represents the electron charge. We notice that this critical current indeed quadratically depends on the tunneling matrix $T$, which reflects the second order perturbation contribution in the tunneling processes.

We next consider the supercurrent carried by the MBS channel. The tight-binding model described by Eq. (1) enters the topological state in the presence of appropriate spin-orbit coupling and large Zeeman energy $V^2_x > \Delta^2 + (\mu-2t)^2$. Four MBSs appears at the four ends of the two chains. The two MBSs in the junction area, $\gamma_1$ and $\gamma_2$ couple together and contribute to the Josephson current. 
They support coherent single electron tunneling, which gives a psudo-$4\pi$ periodicity in the current-phase relation (see appendix for details). We obtain a Josephson relation $I = I_m \langle i \gamma_1 \gamma_2 \rangle  \sin \frac{\theta}{2} $ with the critical current,
\begin{eqnarray}\label{Majoranacritical}
I_2 = \frac{e \nu  T}{2\hbar},
\end{eqnarray}
where $\nu = \sum_\sigma \langle  \gamma_1 \gamma_2  c_{L, 0 ,\sigma}^{\dagger} c_{R,0,\sigma}\rangle$ is the overlapping between the wave function created by the tunneling Hamiltonian and the wave function created by the two MBSs. The critical current is linearly depending on the tunneling matrix, which is a signature of the degenerate perturbation contribution. 
The current-phase relation of this Josephson current is unique. It not only depends on the phase difference $\theta$ but also depends on the quantum average Majorana parity state $\langle i \gamma_1 \gamma_2 \rangle$. This gives an extra degree of freedom for the Josephson relation, which leads to three typical scenarios. First, if the quasiparticle poisoning and the Majorana poisoning are totally ignored, the Majorana parity operator $i \gamma_1 \gamma_2$ is a conserved quantity. Then we have an exact $4\pi$ period Josephson relation $I = \pm I_1 \sin \theta /2$. This is the so-called fractional Josephson effect which has been well discussed in literature\cite{beenakkerprb,law2}. 
Second, if the adiabatic processes are considered and all poisoning are included\cite{kitaevwire}, the Josephson relation will reduce to a skewed $2\pi$ period one $I \approx I_2 \frac{\cos \theta /2}{|\cos \theta / 2|}\sin \theta /2$. In experiments, this skewed $2\pi$ period Josephson relation is also used as a signal for MBSs\cite{moler,lv}.
Finally, if the Majorana parity operator is treated as a quantum psudo-spin, we would have a correlated dynamics of the phase difference and the psudo-spin\cite{wangLZS}. The former obeys the classical Newton equation and the latter obeys the Shcr\"{o}dinger equation. This correlated dynamics leads to rich phenomena, and provides methods for controlling the Majorana qubit\cite{wangLZS}.

We focus on the comparison between the critical currents of the Majorana channel and the conventional channel, as presented in Eqs (3) and (4). The Majorana channel gives a Josephson current which linearly depends on the tunneling matrix $T$, while the conventional Josephson current is a quadratic function of $T$. We look at the ratio between these two critical currents, and obtain a ratio of
\begin{eqnarray}
R = I_2 / I_1 = \frac{  (1 - \mu^2/ 4 t^2)  \nu t^2   }{  \Delta T}.
\end{eqnarray}
We find that this ratio $R$ must be a large number for tunneling barrier junction which has a small tunneling matrix. It becomes larger when we reduce the tunneling matrix $T$ by increasing the voltage on the gate. In principle, we have no limit on reducing the tunneling matrix. At small $T$ limit, we face a tremendous increasing of the critical current when the system goes from the trivial phase into the topological phase. If we check the behavior of the critical current as a function of the control parameter, say Zeeman energy, we must find a sharp step-like function at the TQPT point.

\section{Critical Current Steps at the Topological Quantum Phase Transition Point}
We have obtained analytic results of the critical current of the junction for two typical parameters. One is for the trivial phase with the critical current shown in Eq. (3) and the other is for the extreme topological phase with the critical current shown in Eq. (4). However, we must access to the critical current for general parameters if we want to study the TQPT, because the TQPT involves a continuous modulation of a control parameter. This task is difficult analytically. Therefore we adopt the numerical BdG approach to obtain Josephson current of the spin-obit coupling chain model shown in Eqs. (1) and (2). In the BdG approach, 
we calculate the energy spectrum $E_n$ of the total Hamiltonian as a function of the phase difference $\theta$, and obtain the Josephson current with phase derivative of the energy spectrum\cite{beenakkerprl},
$I(\theta)=\frac{e}{\hbar} \sum_n {\partial E_n(\theta)}/ {\partial \theta}$.
The maximal value of the Josephson current gives the critical current of the junction. We change the Zeeman energy $V_x$ and calculate the critical current every time. Finally, we obtain the critical current as a function of the Zeeman energy, and show the results in Fig. 2. We study three different tunneling matrix. For clear comparison, we normalized each curve with its value for $V_x =0$. We see that all three critical currents exhibit step-like features around the TQPT point $V_x = \sqrt{\Delta^2 + (\mu-2t)^2}$. The steps become steeper for smaller tunneling matrix due to the increasing of the ratio $R$ in Eq. (5). 
This anomalous critical current provides a clear marker for the TQPT. We already show that the critical current enhancement directly comes from the Majorana channel. Therefore, the critical current anomaly also gives a valid signal for the existence of MBSs. 

\begin{figure}[h]
\begin{centering}
\includegraphics[clip=true,width=0.5\columnwidth]{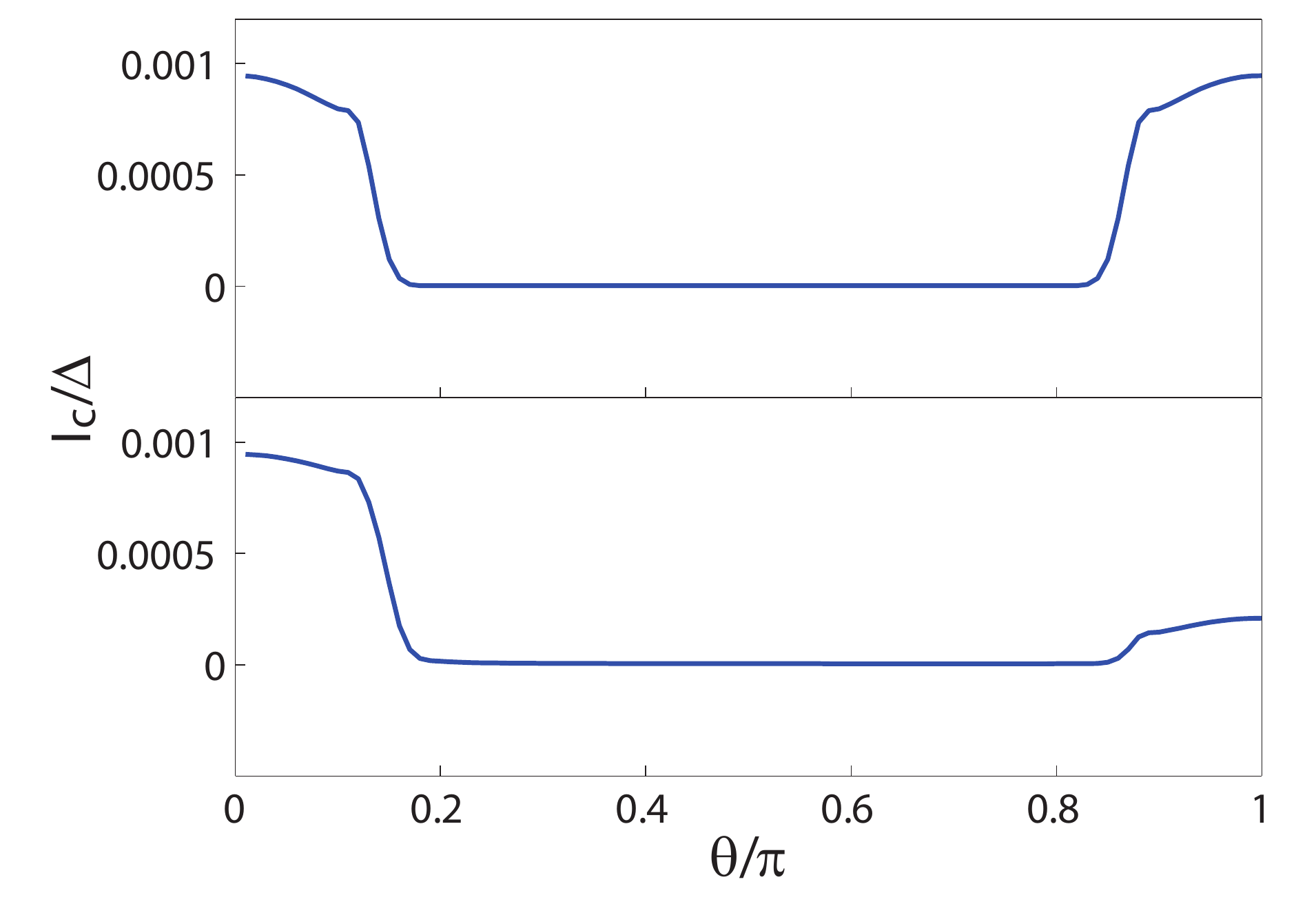}
\caption{(Color online). The critical current of the one-band Josephson junction in response to a rotating Zeeman field (a) on the entire junction and (b) only on the right lead.}
\end{centering}
\label{F4}
\end{figure}

In experiments, another method to achieve TQPT is to rotate the Zeeman field\cite{mourik}. The study of rotating the Zeeman field on nanowire systems has provided interesting results. Zero energy excitations are found robust within certain Zeemn field directions, which agree with the theoretical predictions. However, the energy gap closing, which should happen at the TQPT points, is not observed\cite{mourik}. Here, we show that the critical current is a good signal for marking the TQPT when the Zeeman field is rotated. The rotation of the Zeeman energy requires a y-component in the Zeeman energy. We add one more term in the Hamiltonian of the chain in Eq. (1),
\begin{eqnarray}
H' = H +  \sum_{\bf {r}, \sigma\sigma^\prime}   c_{\bf {r},\sigma}^\dag (V_y\sigma_y)_{\sigma\sigma^\prime}
c_{\bf {r},\sigma'},
\end{eqnarray}
where $V_y$ represents the Zeeman energy in the y-direction. We fix the amplitude of the Zeeman energy $V = \sqrt{V^2_x + V^2_y }$ and rotate its angle. Then we calculate the critical current of the junction. We first consider a rotation of the magnetic field in the total junction, and show the results in Fig. 3a. We find that 
The critical current is at the maximum when the direction of the Zeeman field is along the x-axis. This large critical current is mainly carried by the Majorana channel. Then the critical current gradually decreases with the rotation of the Zeeman energy.This should come from the reduction of the MBS wave function overlapping factor $\nu$. When the rotation becomes larger, the TQPT happens and the critical current suddenly drops to a small value. This small critical current is entirely carried by the quasiparticle channel, therefore is not sensitive to the rotation of the Zeeman field. Finally, the Zeeman field is rotated in the reverse direction and the TQPT happens again. The critical current rises and reaches the maximum when the magnetic field rotates to the inverse of the x-axis. We then consider the situation for only rotating the Zeeman field on the right chain, with the results shown in Fig. 3b. We obtain similar results. However, the critical current is asymmetric to the rotation angle, since the MBS wave function overlapping is suppressed when the two chains have opposite Zeeman fields.

The angle dependence of the critical current gives a full landscape. This should be helpful for the experiments. The critical current is easy to measure. Therefore, it could provides information on the TQPT for those systems where the energy gap closing is not observed at the TQPT points.

\section{Multi-band Model}
Realistic nanowires can be multiband systems, with the number of the sub-bands determined by the width of the wire. We model the multi-band wires by expanding the chain into multi-layers. Therefore, the Hamiltonian in Eq. (1) should be slightly modulated into,
\begin{eqnarray}
H_{L,R}&&= 
- t  \sum_{ \langle \bf {r } , \bf {r'} \rangle, \sigma} c_{\bf {r} , \sigma}^\dag c_{\bf{r'}\sigma}
+ \sum_{\bf {r}, \sigma\sigma^\prime}   c_{\bf {r},\sigma}^\dag (V_x\sigma_x)_{\sigma\sigma^\prime}
c_{\bf {r},\sigma'}
\\\nonumber
&& +  \frac{\eta}{2}  \sum_{\bf{r}, \sigma, \sigma'}   (c_{\bf{r+\delta_x}, \sigma}^{\dag}   (i\sigma_y)_{\sigma\sigma^\prime}   c_{\bf {r},\sigma'} -c_{\bf{r+\delta_y}, \sigma}^{\dag}   (i\sigma_x)_{\sigma\sigma^\prime}   c_{\bf {r},\sigma'} )
\\\nonumber
&& +\sum_{\bf{r}} (\Delta e^{i\phi_{L,R}}  c_{\bf {r},\uparrow}^\dag c_{\bf {r},\downarrow}^\dag+h.c)
- \mu\sum_{  \bf {r } , \sigma} c_{\bf {r} , \sigma}^\dag c_{\bf{r}\sigma},
\end{eqnarray}
where $\bf{r}$ is the position of the multi-layer chain, and $\bf{\delta_x}$ is the unit step in the x-direction.
The energy spectrum and eigenfunctions of this multi-chain model have been studied\cite{potter}. The results show that end MBSs fuse and disappear for even number of layers, while MBSs still exist for odd-layer systems. Here we study the critical current of the multi-chain system by adding a tunneling Hamiltonian similar to Eq. (2), where the electron tunneling are restricted to the same layer. We show the critical current for one, two, and three layers in Fig. 3. 
We see that the critical current anomaly exist for odd number of layers and disappear for even number of layers. This directly links the critical current anomaly with the existence of free MBSs. For even-layer system, all MBSs pair together and forms local Andreev bound states. These Andreev bound states also carry Josephson current, but through second order tunneling processes. Therefore, the critical current enhancement disappears.
We also notice that the step become smoother with increasing layers. It comes from the enhancement of the critical current of quasiparticle channels, which increases with increasing layers. If the number of layers is large, the system changes from quasi-1d into quasi-2d. Then the critical current enhancement will entirely disappear.

\begin{figure}[h]
\begin{centering}
\includegraphics[clip=true,width=0.5\columnwidth]{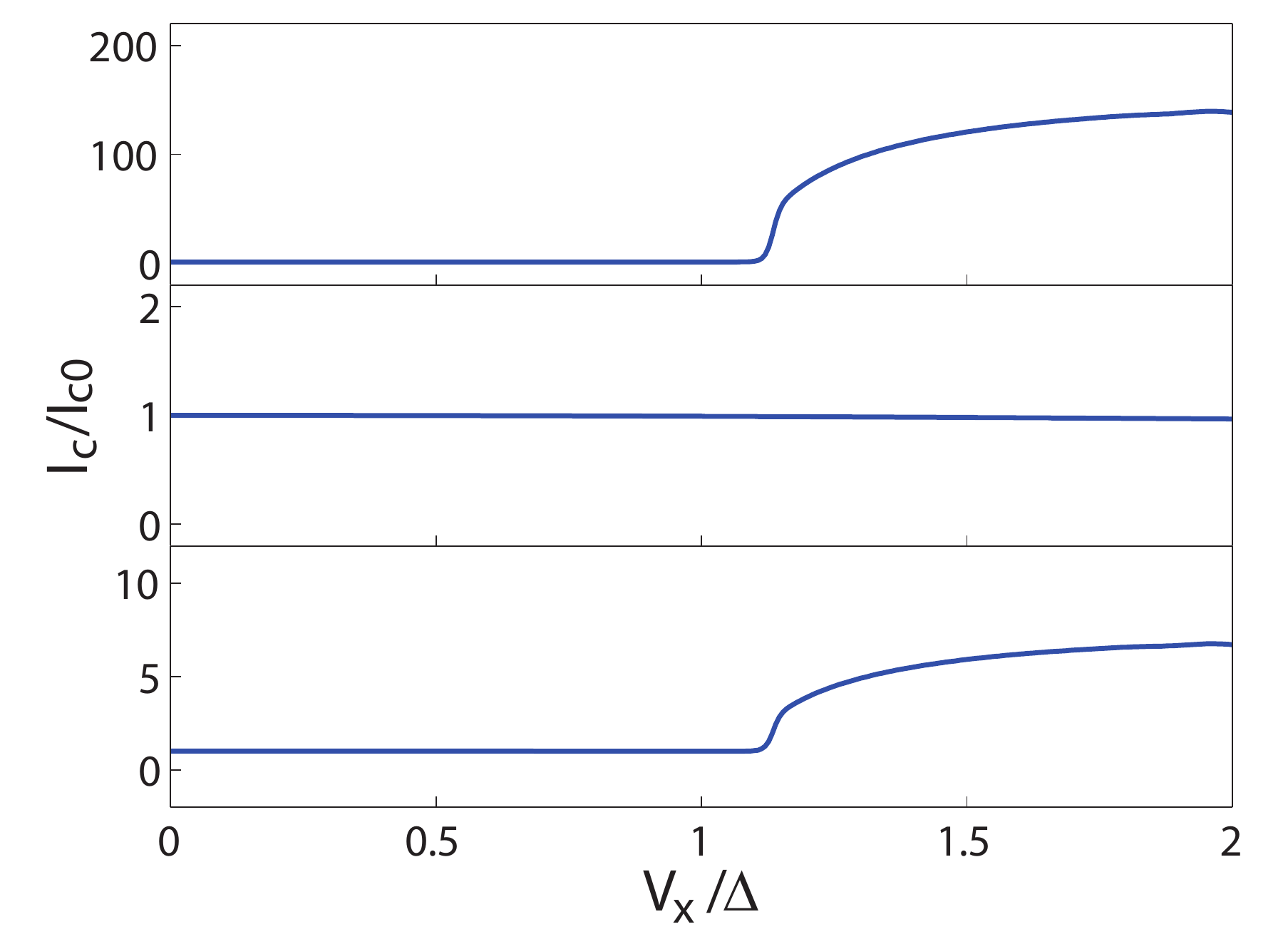}
\caption{(Color online). The critical current of the multi-band Josephson junction in response to the Zeeman energy $V_x$ for the case of (a) one-band system, (b) two-band system, and (c) three-band system. Parameters are taken the same as in Fig. 2. Each current value in $V_x$=0 are $1.6\times10^{-5}\frac{e\Delta}{2\hbar}$, $2.6\times10^{-4}\frac{e\Delta}{2\hbar}$, $3.4\times10^{-4}\frac{e\Delta}{2\hbar}$, respectively.}
\end{centering}
\end{figure}

\section{Conclusion}

In summary, we study the critical current in the topological nanowire Josephson junction, where a potential barrier suppresses the tunneling strength. We investigate the tunneling processes which contribute to the Josephson current and give analytical results of the critical current for typical parameters in the topological phase and the trivial phase. We find that the ratio between these two critical currents is linearly depending on the tunneling matrix. Therefor there must be a critical current anomaly between these two phases when the tunneling matrix is small enough. We then use Bogoliubov-de Gennes approach to investigate the critical current under increasing and rotating Zeeman fields. We show that the critical current indeed has a sharp step-like feature at the topological quantum phase transition. We then study  
multi-band systems, and find that the critical current enhancement disappears for even-layer systems, where the Majorana bound states fuse into Andreev bound states. Our study provides an experimentally accessible signal for the topological quantum phase transition and the existence of Majorana bound states.

\section*{Acknowledgement}
This work is supported by NSFC-11304400, NSFC-61471401, and SRFDP-20130171120015. Q.F.L is supported by NSFC-11574215. D.X.Y. is supported by NSFC-11074310, NSFC-11275279, SRFDP-20110171110026, and NBRPC-2012CB821400.

\appendix

\section{Critical Current from Quasiparticle Channel}
The Josephson current from the quasiparticle channel can be obtained by considering the trivial limit of zero spin-orbit coupling and Zeeman energy, $\eta = V_x = 0$. For this simple parameter, the two superconducting segments become conventional s-wave superconductors, and the tunneling Josephson current is given as\cite{Mahan},
 \begin{eqnarray}
I(t)&&= -e  \langle \frac{dN_L}{dt} \rangle   =- \frac{ei}{\hbar}\langle  [ H, {  N_L}]\rangle
\\\nonumber
&&=    - \frac{ei}{\hbar} \sum_\sigma  \langle \psi(t)| T  c^{\dagger}_{L,0,\sigma} c_{R,0,\sigma} -h.c |\psi(t)\rangle,
\end{eqnarray}
where where $|\psi(t)\rangle$ is the ground state wave function of the total system.
We implement periodic boundary conditions for the two lead of the junction. This does not change the results since no edge state exist in the trivial phase. Then we can make Fourier transformations $c^\dagger_{L,j,\sigma} = \frac{1}{\sqrt {N}}\sum_k e^{-ikj} c^\dagger_{k,\sigma}$, and  $c_{R,j,\sigma} = \frac{1}{\sqrt {N}}\sum_p e^{ipj} c_{p,\sigma}$, with $N$ the number of sites at each lead. The current is rewritten as,
\begin{eqnarray}
I(t ) && =   - \frac{e}{N \hbar}  {\rm Im} \sum_{{ k},{p},\sigma}  \langle\psi(t)|  T c^{\dagger}_{{k},\sigma} c_{{ p},\sigma}  |\psi(t) \rangle
\\\nonumber
&& =    \frac{e}{N \hbar}  {\rm Im} \sum_{{ k},{p},\sigma}  T  \langle\psi_0| e^{-iHt}    c^{\dagger}_{{k},\sigma} c_{{ p},\sigma} e^{iHt}  |\psi_0\rangle,
\end{eqnarray}
where $|\psi_0 \rangle$ is the ground state at the time origin. For {\it non-degenerate} system, this ground state evolution is obtained by going to the {\it interaction representation},
\begin{eqnarray}
I_S(t )
=   \frac{e}{N\hbar}  {\rm Im}  \sum_{{k},{p},\sigma}  T   \langle \phi_0 | S^\dagger(t,-\infty) \hat c^{\dagger}_{{k},\sigma} (t) \hat c_{{ p},\sigma}(t) S(t,-\infty) |\phi_0 \rangle, 
\end{eqnarray}
where $\hat c$ represents the operators in the interaction representation, $\phi_0$ is the ground state wave-function in absence of the tunneling Hamiltonian, $S(t,t')$ is the S-matrix obtained by the evolution operator, which could be expanded to the first order as,
\begin{eqnarray}
S(t,-\infty)=\left[1-i\int_{-\infty}^t dt_1 \hat H_T(t_1)\right] + O(T)^2,
\end{eqnarray}
We omit higher order perturbations and take the lowest order contributions. The current is expressed as a correlation function,
\begin{eqnarray}
I(t)= - \frac{eT^2}{N^2 \hbar}  \sum _{k,p,k',p'\sigma}\int_{-\infty}^{\infty} dt' \Theta(t-t')  \langle \phi_0|  [\hat c^{\dagger}_{{k},\sigma} (t) \hat c_{{ p},\sigma}(t),\hat c^{\dagger}_{{k'},\sigma} (t') \hat c_{{ p'},\sigma}(t')]|\phi_0 \rangle. \nonumber\\
\end{eqnarray}
This correlation function can be analytically calculated with the standard {\it Green function} technique\cite{Mahan}, where we could draw out the Josephson current part as,
\begin{eqnarray}
I = 2 e {\rm Im } [ e^{-2ieVt/\hbar} \Pi_{ret} (eV)],
\end{eqnarray}
where the retarded Green function is the analytic continuation of the Matsubara Green function,
\begin{eqnarray}
\Pi_{ret} (i\Omega) = 2 T^2 \sum_{k,p,i\omega} \Im^\dagger(k,i\omega) \Im(p, i\omega-i\Omega), 
\end{eqnarray}
where $\Im^\dagger$ is the off-diagonal Matsubara Green function in superconductors. We finally obtain the dc Josephson current by taking $eV = 0$,
\begin{eqnarray}
I = I_1 \sin \theta,
\end{eqnarray}
with a critical current,
\begin{eqnarray}
I_1 =   \frac{e \Delta^2  T^2}{ N^2 \hbar}   \sum_{k,p} \frac{1}{E_{ k} E_{ p} (E_{ k}+E_{ p})},
\end{eqnarray}
where the energy spectrum of the superconductor is $E_{k} = \sqrt{(-2t \cos k - \mu)^2 + \Delta^2}$. The summation for the critical current changes into an integration in the continuous limit,
\begin{eqnarray}
I_1 =\frac{ 4e \Delta^2 T^2 N_L N_R }{N^2 \hbar}\int^\infty_\Delta dE \frac{\rho(E)}{E}  \int^\infty_\Delta dE' \frac{\rho(E')}{E'} \frac{1}{E+E'},
\end{eqnarray}
where $N_L$ and $N_R$ are the average density of states near the Fermi surface, and $\rho(E) = \frac{E}{\sqrt{E^2 - \Delta^2}}$ is the superconducting density of states.
This integral over energy is an elliptic integral, which can be integrated out as,
\begin{eqnarray}
I_1 && = \frac{2 \pi^2 e N_L N_R \Delta T^2}{N^2 \hbar}.
 \end{eqnarray}
For one-dimensional tight-binding system, the Fermi wave vector is determined by the hopping term and the chemical potential $k_F = \arccos (- \mu/ 2t)$. The density of the states at the Fermi surface is the inverse of the slope of the dispersion function $N_L = N_R = \frac{N}   {\pi 2  t \sin k_F}  = N / 2 \pi t \sqrt{1 - \mu^2 / 4 t^2}$. Plugging this back into Eq. (A9), we arrive at the formula for the critical current in Eq. (3),
\begin{eqnarray}
I_1 =  \frac{  e  \Delta T^2}{  2 (1 - \mu^2/ 4 t^2) \hbar t^2}.
\end{eqnarray}

\section{Josephson Current From Majorana Channel}
When the junction enters the topological phase, four isolated MBSs appears in the edges of the wire. Two of them $\gamma'_L$ and $\gamma'_R$ locate at the ends of the wire, while the other two $\gamma_L$ and $\gamma_R$ locate at the two sides of the junction. These four MBSs form the four-fold degenerate ground state for the topological superconductor. This ground state degeneracy prevents the application of  the standard interaction picture and S-matrix expansion. However, we can use a degenerate perturbation approach to obtain the current. We illustrate the calculation of the Majorana Josephson current with the simple Kitaev model which grasps the essence of the topological superconductivity. The Hamiltonian for the two segments of the wire writes as, 
\begin{eqnarray}
H_{{\rm K}} = && \sum_{j} \left[- t c^\dagger_{j} c_{j+1} + e^{i\phi_{L,R}} c_{j} c_{j+1}  + h.c. \right]  - \mu \sum_j c^\dagger_{j} c_{j},
\end{eqnarray}
where $\phi_{L,R}$ represents the superconducting phase for the left and right superconductors, respectively. We take identical parameters for the two superconductors for simplicity.
Follow Kitaev,
we take the transformation from electron representation to Majorana representation\cite{kitaevwire},
\begin{eqnarray}
&& \gamma_{j,A} = e^{i \phi_{L,R}} c_j + e^{-i \phi_{L,R}} c^\dagger_j
\\\nonumber
&&  \gamma_{j,B} = -i e^{i \phi_{L,R}} c_j + ie^{-i \phi_{L,R}} c^\dagger_j.
\end{eqnarray}
Then the model is rewritten with Majorana operators,

\begin{eqnarray}
H_{{\rm K}} 
 = && \frac{ - i {\mu} }{2}\sum_{j}  \gamma_{j,A } \gamma_{j,B} + \frac{i(t+ \Delta)}{2}\sum_{j}   \gamma_{j,B} \gamma_{j+1,A}
\nonumber\\
 &&- \frac{i(t- \Delta)}{2}\sum_{j}  \gamma_{j,A} \gamma_{j+1,B}  ].
\end{eqnarray} 

The electron tunneling across the junction connects the left and the right superconductors, which is described by a tunneling Ham
\begin{eqnarray}
 H_{TK} =  T  c^{\dagger}_{L,0} c_{R,0} + h.c.
\end{eqnarray}
where $T$ is the tunneling matrix.
The current through the junction comes from the electron tunneling, which can be expressed as\cite{Mahan},
 \begin{eqnarray}
I(t) =  - \frac{ieT}{\hbar} \langle \psi(t)|   c^{\dagger}_{L,0} c_{R,0} -h.c |\psi(t) \rangle,
\end{eqnarray}
We notice that the expression for the current is quite similar to the conventional junction. However, we have a key difference that the ground state $|\psi(t) \rangle$ is now degenerate. Therefore it is impossible to go to the interaction representation and apply Green function technique. We must calculate the current with the  degenerate perturbation theory. In the zero's order degenerate perturbation approach, we restrict the wave function $|\psi(t)\rangle$ into the Hilbert-subspace expanded by the degenerate ground states of the unperturbed Hamiltonian without tunneling $H_{KT}$, and ignore all terms which project the wave function out of this subspace. We remember that the degenerate ground states of Kitaev model are defined by the end Majorana operators, therefore, we massage the formula Eq. (B5) by expanding the electron operators with Majorana operators,
\begin{eqnarray}
I (t) = &&- \frac{eT i}{2\hbar} \langle \psi(t)|(\gamma_{L,0,B}\gamma_{R,0,A} -\gamma_{L,0,A}\gamma_{R,0,B} ) \sin\frac{\theta}{2} 
\nonumber\\
&&+  (\gamma_{L,0,A}\gamma_{R,0,A}+\gamma_{L,0,B}\gamma_{R,0,B} ) \cos \frac{\theta}{2} |\psi(t) \rangle,
\end{eqnarray}
where $\theta = \phi_L - \phi_R$ is the phase difference across the junction.
Let us first consider the special  parameter of $\mu= 0$ and $t=\Delta$, where the two zero energy Majorana operators $ \gamma_{L,0,B}$ and $\gamma_{R,0,A}$ are the Majorana zero modes which define the unperturbed degenerate ground states. Then we drop all  three terms which project $|\psi \rangle$ out of the degenerate ground state subspace, and take the only term which is expressed by the zero energy Majorana operators,
 \begin{equation}
I_M(t)=
  \frac {eT} {2\hbar}  \sin({\theta}/{2})  \langle \psi(t)| -i\gamma_{L,0,B}\gamma_{R,0,A} |\psi(t)\rangle,
\end{equation}
where the ground state wave function $|\psi (t)\rangle$ evolves according to the Schr\"{o}dinger equation,
\begin{equation}
-i \hbar \frac {d} {dt} |\psi(t)\rangle =\left[\frac{T}{2} (-i\gamma_{L,0}\gamma_{R,0})  \cos ({\theta}/{2}) \right] |\psi(t) \rangle.
\end{equation}

For the general parameters of $\mu \neq 0$ or $t \neq \Delta$, the zero energy MBSs are the combination of the Majorana operators\cite{kitaevwire},
\begin{eqnarray}
\gamma_{L} = \sum_j a_j  \gamma_{L,j,B},
\gamma_{R} = \sum_j b_j \gamma_{R,j,A}.
\end{eqnarray}
These two MBS are localized near the junction area, that is, $a_0 \approx 1, a_{j \neq 0} \approx 0$, and same for $b_j$. Then the current is expressed as,
 \begin{eqnarray}
I_M(t) =
 \frac { \nu eT} {2 \hbar}     \sin({\theta}/{2})  \langle \psi(t)| -i\gamma_{L}\gamma_{R} |\psi(t)\rangle,
\end{eqnarray}
where $\nu = a_0 b_0 /4$ comes from the overlapping between the tunneling operator $c^{\dagger}_{L,0} c_{R,0}$ and the MBS operator $i \gamma_L \gamma_R$. We arrive at the critical current shown in Eq. (4),
\begin{eqnarray}
I_2 = \frac{e \nu  T}{2\hbar}.
\end{eqnarray}

For these general parameters, there is a exponentially small but non-zero coupling between the two MBSs at the junction $\gamma_L$ and $\gamma_R$ and the two MBSs at the two ends $\gamma'_L$ and $\gamma'_R$, which provides a small coupling Hamiltonian,
\begin{eqnarray}
H_\delta = i\delta_L\gamma'_L\gamma_L + i\delta_R \gamma_R \gamma'_R.
\end{eqnarray}
where $\delta_{L,R}$ are exponentially suppressed by the length of the wire. This coupling Hamiltonian is small; however, it qualitative changes the quantum dynamics of the ground state wave function $\psi(t)$, which is now governed by the Schr\"{o}dinger equation,
\begin{equation}
-i \hbar \frac {d} {dt} |\psi(t)\rangle =\left[ \frac{- i \nu T}{2}  \gamma_{L}\gamma_{R}  \cos ({\theta}/{2}) + H_\delta\right] |\psi(t) \rangle.
\end{equation}
We see that the $H_\delta$ break the local parity conservation defined by the two MBSs $\gamma_L$ and $\gamma_R$ around the junction, thereby in principle destroys the $4\pi$ period Josephson effect. 

\section{BdG Formalism for Josephson Current}
We use BdG equation to get the Josephson current by solving Hamiltonian,
\begin{eqnarray}
H=H_L+H_R+H_T,
\end{eqnarray}
where $H_L$, $H_R$ and $H_T$ are described in euqtion (1) and (2). Each term in Hamiltonian is bilinear term of $c^{\dag}$ and $c$, with a series of parameters such as t, $\Delta$ and $\theta$. For a series of constant parameters such as $t_0$, $\Delta_0$ and $\theta_0$, we can use BdG method to transform electrons and holes into quasi-particles which can diagonalize the Hamiltonian. The relation between electrons, holes and quasi-particles is BdG equation,
\begin{eqnarray}
c_{i\uparrow}= \sum_n u_{n\uparrow}\gamma_n+v_{n\uparrow}^\ast\gamma_n^\dag\\
c_{i\downarrow}= \sum_n u_{n\downarrow}\gamma_n+v_{n\downarrow}^\ast\gamma_n^\dag,
\end{eqnarray}
where $\gamma_n^\dag$($\gamma_n$) represents the creation(annihilation) operator of quasi-particles, $u_n$ and $v_n$ are a series of parameters that we choose to diagonalize the Hamiltonian. Then the Hamiltonian is written as
\begin{eqnarray}
H_{BdG}=E_g+\sum_n \epsilon_n\gamma_n^\dag\gamma_n,
\end{eqnarray}
where $E_g$ represents the ground-state energy, and $\epsilon_n$ is the quasi-particle energy. All the negative quasi-particle wave function are summed up to forms the ground state of the Hamiltonian.
The quasi-particle energies $\epsilon_n$ are affected by the value of a series of parameters. To calculate the current-phase relation, we choose phase difference $\theta$ to be changeable. In this way, the quasi-particle energies $\epsilon_n$ are functions of phase difference $\theta$, and the Hamiltonian is the function of phase difference $\theta$ too,
\begin{eqnarray}
H_{BdG}(\theta)=E_g+\sum_n\epsilon_n(\theta)\gamma_n^\dag\gamma_n.
\end{eqnarray}
Then we take the derivative of all the negative quasi-particle energies to give the current phase relation,
\begin{eqnarray}
I(\theta)=\frac{ge}{\hbar}\sum_n\frac{\mathrm{d}}{\mathrm{d}\theta}\epsilon_n(\theta),
\end{eqnarray}
where $g$ represents the factor that counts spin and other degeneracies. To simplify the calculation, we make $\frac{ge}{\hbar}=1$.




\bibliographystyle{elsarticle-num}
\bibliography{<your-bib-database>}



\end{document}